\documentclass[
    ,final            
  ]
  {aipproc}

\layoutstyle{6x9}

\def\ltsima{$\; \buildrel < \over \sim \;$}
\def\lsim{\lower.5ex\hbox{\ltsima}}
\def\gtsima{$\; \buildrel > \over \sim \;$}
\def\gsim{\lower.5ex\hbox{\gtsima}}
\begin{document}

\title{Linear polarization on Gamma-Ray Bursts: from the prompt 
to the late afterglow}

\author{Davide Lazzati}{
  address={Institute of Astronomy, University of Cambridge, Madingley
Road, CB3 0HA, Cambridge, UK} }

\begin{abstract}
The past year has witnessed a large increase in our knowledge of the
polarization properties of Gamma-Ray Burst (GRB) radiation. In the
prompt phase, the measurement (albeit highly debated) of a large
degree of linear polarization in GRB~021206 has stimulated a deep
theoretical study of polarization from GRB jets. The optical afterglow
of GRB~030329, on the other hand, has been followed thoroughly in
polarimetric mode, allowing for an unprecedented sampling of its
polarization curve. I will review the present status of theories and
observations of polarization in GRBs, focusing on how polarimetric
observations and their modelling can give us informations on the
structure and magnetisation o GRB jets which is not possible to obtain
from their light curve.
\end{abstract}

\maketitle


\section{Introduction}

Linear polarization has revealed to be a characteristics of GRBs
throughout their entire evolution. The recent claim by Coburn \&
Boggs\cite{CB03} that the prompt emission of GRB~021206 was polarized
at a very high level has stimulated a thorough analysis of the
polarizing properties of the jet geometries in GRB outflows, drawing
attention on the possibility that magnetic fields may be advected from
the central source rather than generated by the internal shock. On the
other hand, observational and theoretical studies of afterglow
polarization has revealed a much more complicated picture than
previously thought, emphasising the importance of the jet structure,
its dynamics and the properties of the ISM in shaping the polarization
curves. Even though direct observations lack, also the optical flash
may be highly polarized, especially if due to a reverse shock in the
burst ejecta rather than to the pair enrichment of the nearby ISM.

In this paper I review the status of theories and observations of
linear polarization in GRBs. The three phases are analysed initially
separately and then their relation discussed. The importance and
insight of polarization is emphasised in all phases.

\section{The prompt phase}

Analysing the scattering geometry of photons in the RHESSI detector,
Coburn \& Boggs \cite{CB03} were able to measure the average linear
polarization of the prompt emission of GRB~021206 in the
[150~keV--2~MeV] energy range. They find that the prompt emission of
the burst is highly polarized, with $\Pi=0.8\pm0.2$. This measurement
was subsequently heavily criticised by Rutledge \& Fox \cite{RF03},
who performed an independent analysis of the same dataset, obtaining a
much smaller number of double-scattered photons and, as a consequence,
merely an upper limit on the linear polarization of the event. Despite
that, the result \cite{CB03} has stimulated a vast theoretical effort
in order to understand under which conditions such a large
polarization could be obtained. Two classes of models have emerged. In
the first class, the origin of polarization is ascribed to the
presence of a large scale ordered magnetic field, which ought to be
advected from the central engine and may play a role in the launching
of the jet itself \cite{Ly03}. In the second class of models, the
magnetic field is supposed to be shock generated and tangled on small
timescales, and the asymmetry required to produce polarization is due
to a particular location of the observer with respect to the jet axis
\cite{W03}. This second class of models can be extended to different
emission mechanisms, such as inverse bulk Compton scattering
\cite{L03a}.

\subsection{Magnetic models}

We define here magnetic models those in which polarization is due to
the large scale geometry of the magnetic field. The magnetic field is
likely to be dominated by a toroidal component, since the radial field
decays faster than the tangential one ($B_r\propto{}r^{-2}$ while
$B_\perp\propto{}r^{-1}$).  One important ingredient of these models
is that the observer, due to the relativistic aberration of photons,
cannot see the whole jet. In fact it is only a small $1/\Gamma$ region
of the jet that is observable and therefore the observer does not
detect the overall toroidal structure of the field (which would wash
out the polarization signal) but a highly ordered patch. This is most
important if the magnetic field is not the dominant component of the
outflow. Since regions of the jet separated by more than $1/\Gamma$
are causally disconnected, it is difficult to envisage a coherent
magnetic field on scales larger than $1/\Gamma$, unless the structure
has been created before the acceleration of the jet (when it was still
connected) and frozen into it. Such a transport seem easier to attain
in a magnetic dominated outflow \cite{P03} and even natural in a
force-free subsonic bubble
\cite{Ly03}.

\begin{figure}
\includegraphics[height=.3\textheight]{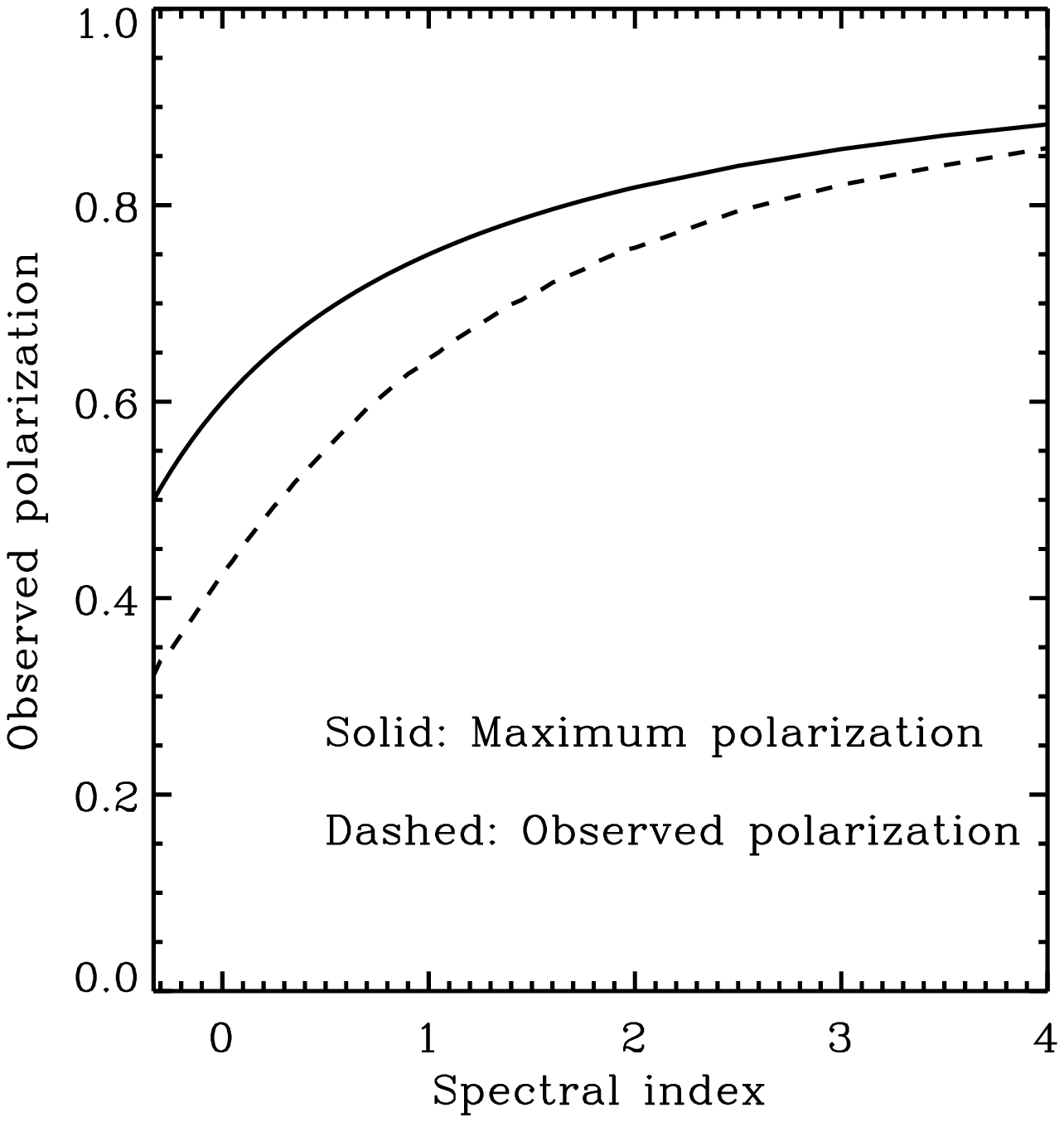}
\includegraphics[height=.3\textheight]{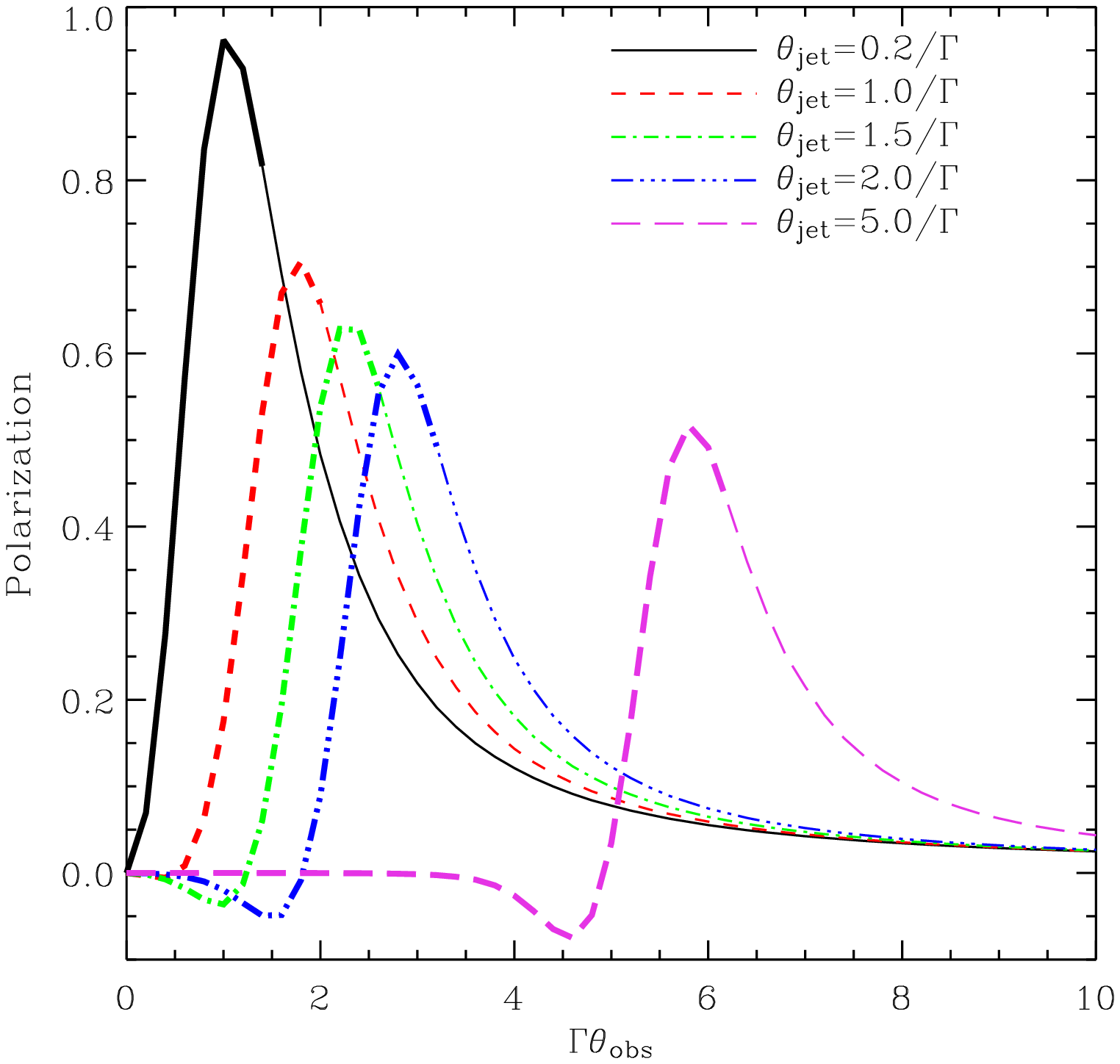}
\caption{{\bf Left panel:}
Maximum synchrotron polarization for a non relativistic uniform field
as a function of the spectral index $\alpha$ (solid line) compared to
the maximum observable polarization from a relativistically moving
uniform field (dashed line). The electron pitch angle distribution is
uniform in both cases..  {\bf Right Panel:} Inverse Compton
polarization as a function of the observing angle $\theta_o$ in units
of $1/\Gamma$ for a uniform jet with sharp edges. Different line
styles show the polarization for jets with different opening
angles. The lines are thicker in the region where the efficiency is
larger than $2.5\%$.  }
\end{figure}

Even if the observer has access only to a fully ordered region of
field, the polarization signal cannot be as large as that
expected from a non relativistic flow. In the classic case:
$\Pi = (p+1)/(p+7/3)$
where $p$ is the power-law index of the electron energy distribution.
The reduction of the observed polarization in the relativistic case is
due to the aberration of photon trajectories. In order to keep the
electric and magnetic field of the wave orthogonal to each other and
to the wave propagation direction, the position angle of polarization
is rotated in different ways as a function of the distance from the
line of sight. After integration, the polarization is reduced by a
factor that depends on the spectral slope of the radiation, and spans
between $10\%$ and $20\%$. The non-relativistic solution is shown with
the relativistic maximum polarization in the left panel of Fig:~1 (see
also \cite{G03}).

A characteristic feature of these models is that any observer located
within the opening angle of the fireball detects a highly polarized
signal, with the exception of those observing the fireball within
$\theta=1/\Gamma$ from the symmetry axis. A level of polarization
comparable to that detected by Coburn \& Boggs cannot be achieved,
even though given the large uncertainties a definitive conclusion
cannot be drawn.

\subsection{Geometric models}

It has been traditionally assumed that the magnetic field responsible
for the synchrotron emission in GRBs is generated at the shock
front. This is a robust conclusion in the afterglow phase, where the
compression of the interstellar field is far too low to produce the
observed radiation. It may hold true also for the prompt phase, in
which case a tangled field would be responsible for the observed
radiation. If this field is tangled in a plane, but compressed in the
direction perpendicular to the plane itself \cite{L80,M99} it is
possible to observe polarized synchrotron radiation since radiation
emitted in the plane, which is maximally polarized in the comoving
frame, is then aberrated toward the observer with an angle
$\theta=1/\Gamma$ \cite{GL99}.

In the afterglow phase, this configuration leads to polarization
of up to several tens of per cent \cite{GL99,S99}, but in the prompt
phase is usually negligible, unless a narrow jet is observed along the
required direction \cite{W03}. Polarization in this context has been
analyzed in several works with synchrotron as the radiation mechanism
\cite{G03,N03} as well as if the photons are
produced by bulk inverse Compton scattering \cite{L03a}. The
difference between the two cases is not of fundamental nature, since
the dependence of polarization on angle is the same for the two
mechanisms. Inverse Compton, on the other hand, can produce larger
polarization since it can be maximally ($100\%$) polarized in the
comoving frame.

The polarization produced by a narrow jet is shown (for the case of
inverse Compton) in the right panel of Fig.~1. In these models a
narrow jet is fundamental since the number of observer that see a
polarized event is limited to those lying in the region between the
edge of the jet and $1/\Gamma$ from it. This region becomes
vanishingly small for $\theta_{\rm{jet}}>10/\Gamma$.  GRB~021206 was
exceptionally bright and, assuming it was at cosmological redshift,
would have had a narrow jet with opening angle of few degrees at most.

To distinguish between magnetic and geometric models is quite easy
once a reasonable number of measurement is available. In the geometric
case only a small fraction of the brightest bursts should be
polarized, while in the magnetic case most of them should.

\section{The flash}

It has been suggested by Granot \& K\"onigl \cite{GK03} that the
reverse shock emission could be as highly polarized as the prompt GRB,
since the plasma responsible for this emission is the same one that
produced the gamma-ray photons. This consideration can be included in
a more general discussion on optical flashes, i.e. bright optical
components that appears at the beginning of the afterglow phase with a
fast decay.

Optical flashes can be produced in two ways: by a reverse shock in a
baryonic jet or by pair enrichment of the external medium in the
vicinity of the GRB \cite{B02}. In most cases the optical flash is
expected to be polarized at a level comparable to the prompt
emission. There is actually only one case in which the optical flash
following a polarized GRB can be unpolarized. In a magnetic model, if
the flash is due to pair enrichment of the ISM, the flash comes from a
shock generated field without the geometric constraints to produce
polarization. In all geometric models the geometry of the prompt phase
is preserved during the flash emission. Finally, if the flash is due
to reverse shock, it should be polarized as discussed by Granot \&
K\"onigl \cite{GK03}.

\section{Afterglow}

The discovery of linear polarization in GRB afterglows dates back to
1999, when a small but highly significant level of polarization was
detected in GRB~990510 \cite{C99}. The detection took place
amid a theoretical effort to predict and/or explain it.

Guzinov \& Waxman \cite{Gr99} discussed the possibility that the shock
generated field organises in coherent patches that expand at a sizable
fraction of the speed of light. They calculated that an observer
should see approximatively $N\gsim50$ patches. If the polarization
inside a patch is $\Pi_0\sim70\%$, the observed one is
$\Pi=\Pi_0/\sqrt{N}\lsim10\%$. Indeed the degree of linear
polarization observed in GRB afterglows is in the per cent range (see
\cite{C03} for a review). However, given the random nature
of the model, the degree and position angle of polarization fluctuates
in time. This is not, at least in some cases, detected in polarization
curves \cite{L03b,Gr03} and for this reason this model is now not
considered particularly promising.

Shortly after the discovery of polarization Ghisellini \& Lazzati
\cite{GL99} and, independently, Sari \cite{S99} proposed a model based on the
assumption that the fireball is beamed in a come and that the shock
generated field is either compressed in the shock plane or elongated
in the axial direction. Polarization is observed if the observer is
not coincident with the cone axis and has a definite and testable
pattern. Polarization at early times is null, increases slowly with
time until it reaches a maximum and then starts to decrease again
until it vanishes. At this moment, which is roughly coincident with
the jet break in the unpolarized lightcurve, the position axis of
polarization rotates by $90^\circ$. Then the polarization curve is
characterised by a second peak, of higher intensity, eventually
vanishing to an unpolarized flux at long times. The intensity of the
polarization signal depends on the off-axis angle: the larger the
off-axis angle the larger the polarization.

\begin{figure}[t]
\includegraphics[height=.3\textheight]{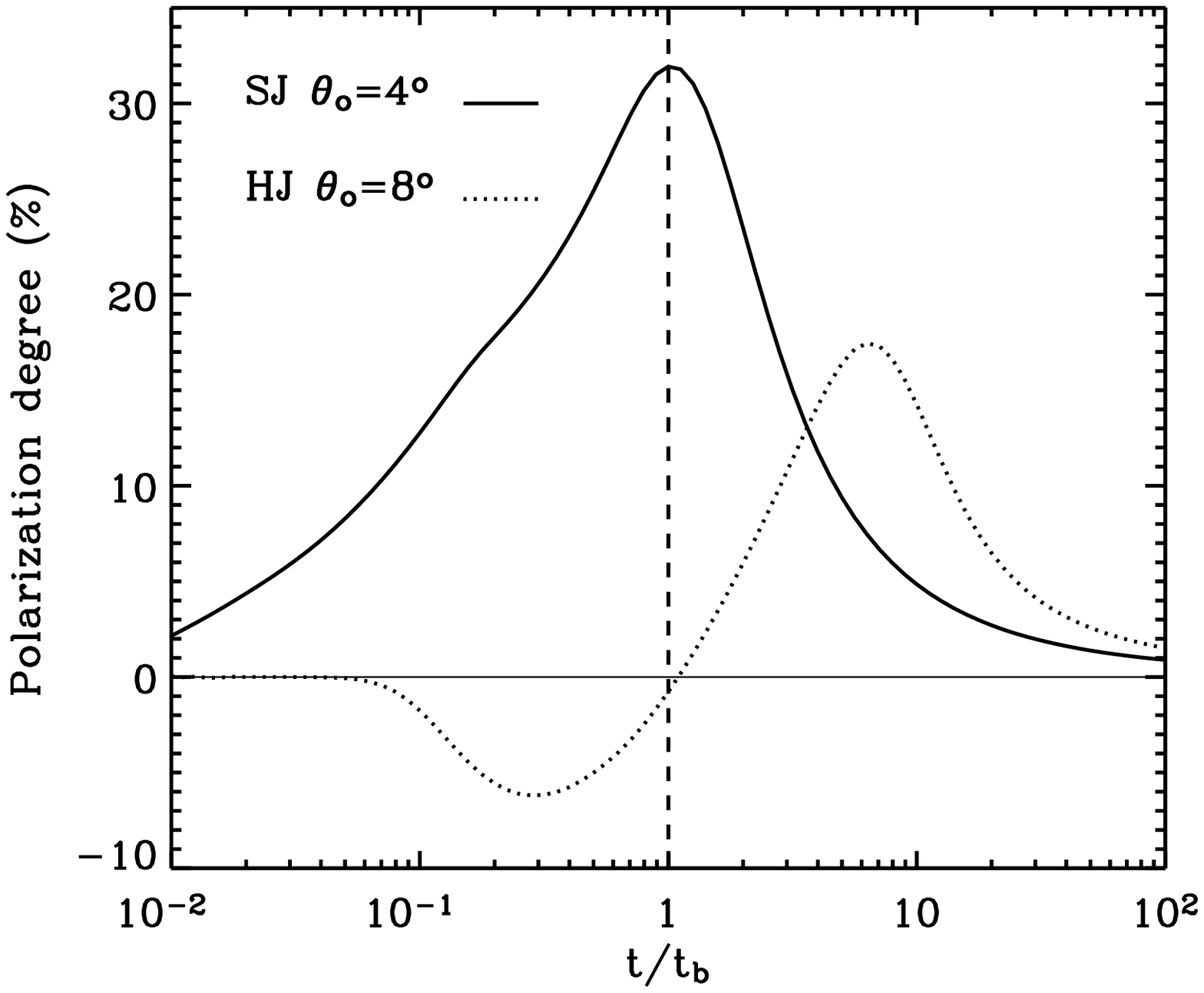}
\caption{Comparison between the polarization curves of a homogeneous 
jet (dotted line) and a structured jet (solid line). The two light
curves are virtually indistinguishable, while polarization behaves in
a markedly different way.}
\end{figure}

These models have been further analyzed and extended by Rossi et
al. \cite{R03}. They studied the effect of different assumptions on
the jet sideways expansion showing that the ratio between the peaks of
polarization is smaller for faster expansion speed. They also
generalised the model to non uniform jets, in which the energy per
unit solid angle decreases as $E_\Omega\propto\theta^{-2}$ where
$\theta$ is the angle with respect to the flow symmetry axis. In this
case the polarization curve is largely different. While it
is null for early and late times, as in the uniform case, it has a
single peak, correspondent in time with the jet break time, and
constant position angle.  In Fig~2 we show the comparison between the
polarization curves for the homogeneous and the structured models. It
is clear that the two curves are different in an easily testable
way. This is particularly important if we consider that the light
curves of the two models are almost indistinguishable. Further
complications to the models have been added by \cite{GK03} who
consider the presence of a coherent component of the magnetic field in
the ISM. The propagation of the polarized light of the OT in the host
and Galactic ISM have been instead discussed by \cite{L03b}.

Comparison of the models with the data has proven difficult. The main
limitation of these models is that they assume that the emissivity of
the fireball is uniform (or strictly dictated by the $\theta^{-2}$
law). Any deviation from this assumption, or inhomogeneity of the
external medium, causes a noise on top of the models in both<
polarization and position angle. Usually this situation is
recognisable in the light curve through the presence of bumps and
wiggles on top of the regular power-law decay. Indeed, every time the
light curve is complex, the polarization curve has a complex
structure, such as in GRB~021004 \cite{Rol03,L03b} and in GRB~030329
\cite{Gr03}. On the other hand, simple polarization curves
seem to be associated to power-law afterglows (GRB020813
\cite{Gor03}).

\section{Comparison of the three phases}

The position angle of polarization should be related in the three
phases. In the intrinsic models (i.e. neglecting polarization induced
by the ISM and by an external magnetic field) the position angle of
the polarization can be either contained in or orthogonal to the plane
containing the jet axis and the line of sight. It is therefore
expected that, should polarization be measured in the future in the
three phases of a single GRB event, the position angle should either
remain constant throughout the whole evolution or rotate by $90^\circ$
between the optical flash and afterglow. It may eventually rotate back
to the original position. Any difference from this simple behaviour
should be considered a sign of an external component in the generation
of polarization.

\section{Summary and conclusions}

The study of polarization evolution in GRBs is highly informative,
albeit difficult. It carries important informations about the jet
structure and dynamics that are hidden in degeneracies of the light
curve, but are emphasised in the polarization curve. Observationally,
the afterglow phase is the most simple to investigate, even though is
may be affected by small-scale inhomogeneities, and constraining the
smoothness of the light curve is of fundamental importance in order
to model a polarization curve. Polarization in the prompt and optical
flash emission is highly informative of the structure of the ejecta,
even though further observations are required in order to establish
the mere existence of polarization in this phases.


\begin{thebibliography}{99}
\bibitem{B02} Beloborodov, A. M., ApJ, 2002, 565, 808
\bibitem{CB03} Coburn, W. \& Boggs, S. E., Nature, 2003, 423, 415
\bibitem{C99} Covino, S. et al., A\&A, 1999, 348, L1
\bibitem{C03} Covino, S., Ghisellini, G., Lazzati, D. \& Malesani, D.,
   2003 (astro-ph/0301608)
\bibitem{GL99} Ghisellini, G. \& Lazzati, D., MNRAS, 1999, 309, L7
\bibitem{Gor03} Gorosabel, J. et al., A\&A subm., 2003 (astro-ph/0309748)
\bibitem{G03} Granot, J., ApJ, 2003, 596, L17
\bibitem{GK03} Granot, J. \& K\"onigl, A., ApJ, 2003, 594, L83
\bibitem{Gr03} Greiner, J. et al., Nature, 2003, 426, 157
\bibitem{Gr99} Gruzinov, A. \& Waxman, E., ApJ, 1999, 511, 852
\bibitem{L80} Laing, R. A., MNRAS, 1980, 193, 439
\bibitem{L03a} Lazzati, D., Rossi, E. M., Ghisellini, G. \& Rees, M. J.,
   MNRAS, 2003, in press (astro-ph/0309038)
\bibitem{L03b} Lazzati, D. et al., A\&A, 2003, 410, 823
\bibitem{Ly03} Lyutikov, M., Pariev, V. I. \& Blandford, R. D., ApJ, 2003, 597, 998
\bibitem{M99} Medvedev, M. V. \& Loeb, A., ApJ, 1999, 526, 697
\bibitem{N03} Nakar, E., Piran, T. \& Granot, J., JCAP, 2003, 10, 5
\bibitem{P03} Proga, D., MacFadyen, A. I., Armitage, P. J. \& Begelman, M. C.,
   ApJ, 2003, 599, L5
\bibitem{Rol03} Rol, E. et al., A\&A, 2003, 405, L27
\bibitem{R03} Rossi, E., Lazzati, D., Ghisellini, G. \& Salomonson, J. D.,
   in prep., 2003
\bibitem{RF03} Rutledge, R. E. \& Fox, D. B., MNRAS subm., 2003 (astro-ph/0310385)
\bibitem{S99} Sari, R., ApJ, 1999, 524, L43
\bibitem{W03} Waxman, E., Nature, 2003, 423, 388
\end{thebibliography}
\end{document}